# Pressure Induced Superconductivity in the New Compound ScZrCo$_{1-\delta}$

Enyu Wang*, Jing Si*, Xiyu Zhu†, Guan-Yu Chen, Hai Lin & Hai-Hu Wen†

Center for Superconducting Physics and Materials, National Laboratory of Solid State Microstructures and Department of Physics, Collaborative Innovation Center for Advanced Microstructures, Nanjing University, Nanjing 210093, China

*These authors contributed equally to this work.

**It is widely perceived that the correlation effect may play an important role in several unconventional superconducting families, such as cuprate, iron-based and heavy-fermion superconductors. The application of high pressure can tune the ground state properties and balance the localization and itineracy of electrons in correlated systems, which may trigger unconventional superconductivity. Moreover, non-centrosymmetric structure may induce the spin triplet pairing which is very rare in nature. Here, we report a new compound ScZrCo$_{1-\delta}$ crystallizing in the Ti$_2$Ni structure with the space group of *FD3-MS* without a spatial inversion center. The resistivity of the material at ambient pressure shows a bad metal and weak semiconducting behavior. Furthermore, specific heat and magnetic susceptibility measurements yield a rather large value of Wilson ratio ~4.47. Both suggest a ground state with correlation effect. By applying pressure, the up-going behavior of**

**resistivity in lowering temperature at ambient pressure is suppressed and gradually it becomes metallic. At a pressure of about 19.5 GPa superconductivity emerges. Up to 36.05 GPa, a superconducting transition at about 3.6 K with a quite high upper critical field is observed. Our discovery here provides a new platform for investigating the relationship between correlation effect and superconductivity.**

## Introduction

The discoveries of superconductivity in cuprate[1-2] and iron-pnictides / chacolgenides[3-5] have revived the enthusiasm of exploring new high temperature superconductors in correlated systems. It is known that both copper and iron in these families are *3d* transition metal elements. In many compounds containing *3d* transition metal elements, the strong interaction of electrons cannot be neglected leading to a correlation effect and subtle balance between localization and itinerancy of *d*-orbital electrons[6]. And this correlation effect induces tremendous and rich physics, including unconventional superconductivity[7], Mott insulating state[8], bad metal, spin density wave, etc.[9,10]. By doping charge carriers or applying external pressure, the correlation effect can be tuned and the ordering states are suppressed, thus triggering possible superconductivity. Following this idea, superconductivity has been found in $Na_xCoO_2 \cdot 1.3H_2O$[11], $MnP$[12], $CrAs$[13], $Ba_{1-x}Na_xTi_2Sb_2O$[14,15], etc.. Therefore, it is a promising way to explore unconventional superconductivity in compounds with *3d* transition metal elements.

As cobalt is a typical *3d* transition metal element and lies between iron and copper,

the superconductors containing cobalt are generally interesting and unusual. Superconductors containing Co include compounds such as $CeCoIn_5$, $Na_xCoO_2 \cdot 1.3H_2O$ and $Zr_2Co$. $CeCoIn_5$ is a well-known unconventional heavy-fermion superconductor[16] with $T_c$ = 2.3 K. And the superconductor $Na_xCoO_2 \cdot 1.3H_2O$ with $T_c \approx$ 4K is fascinating for its strong electron correlation effect, which may have a *d*-wave gap symmetry reflecting unconventional superconducting pairing mechanisms[17]. In the cobaltite $Na_{0.7}CoO_2$, a Curie-Weiss metallic phase was reported, which was argued to be induced by correlation effect[18, 19]. The cobalt alloy $Zr_2Co$, which crystalizes in the $CuAl_2$ structure (space group: *I4/MCM*) was also reported as a superconductor with $T_c$ = 5 K[20]. And the NMR investigation revealed that it could be an itinerant antiferromagnet in the normal state[21]. Here, we report the discovery of a new compound $ScZrCo_{1-\delta}$ with a non-centrosymmetric structure. It does not crystalize in the tetragonal $CuAl_2$ structure but in the cubic $Ti_2Ni$ structure with the space group *FD3-MS*[22]. An anomalous ground state is found and superconductivity is induced by pressurizing the material to a high pressure.

**Results**

**Characterization of the crystal structure and composition.** Fig. 1a shows the X-ray diffraction patterns for the sample $ScZrCo_{1-\delta}$ and the Rietveld fitting curve (the red line) to the data[23]. The Rietveld refinement gives a quite good agreement between the measured data and the calculated profile with the agreement factor $R_{WP}$ = 1.551 %. And no obvious impurity peaks are found. The lattice parameters for the cubic unit cell are determined to be $a = b = c = 12.22$ Å. The inset shows the schematic structure of ScZrCo$_{1-\delta}$ with the space group $FD3\text{-}MS$. And as shown in Table 1, there are two sites for Sc and Zr, which are occupied by Sc and Zr atoms in approximately the 1:1 ratio. And the occupancies of each site for Sc and Zr are given in Table 1, this is obtained by the best Rietveld fit ($R_{WP}$ = 1.551%). Apparently, all Co atoms occupy the same site, and every four nearest Co atoms form a tetrahedron, with a Co-Co bond with length of 3.043 Å. This tetrahedron is surrounded by Sc and Zr atoms. Every tetrahedron together with other four nearest tetrahedron assembles a larger tetrahedron. Thus, there is no spatial inversion symmetry in this compound. Fig. 1b shows a representative energy dispersive X-ray spectrum (EDS) of a micro area on our ScZrCo$_{1-\delta}$ polycrystalline sample. The SEM photograph of the sample is shown in the right upper position of Fig. 1b, while the element ratio in this area is very close to 1: 1: 1. To estimate the accurate ratio of the compound, we have detected the compositions in eight areas on three samples. They are quite close to each other. The averaged ratio among the compositions Sc: Zr: Co is 1.00: 1.09: 0.91. It seems that the Co has a slight deficiency, which may be substituted by the extra Zr.

**Magnetic susceptibility and specific heat.** The temperature dependence of magnetic susceptibility is shown in Fig. 2a with an applied magnetic field of 1 T. There is a small hump at about 50 K, which is caused by the antiferromagnetic transition of frozen oxygen. Since this phenomenon appears quite often in our SQUID and is irrelevant to the related physics here, we have already removed this hump by using a wave packet of Gaussian function. Using the Curie-Weiss law, we fit the data by the equation $\chi(T)$ = $c/(T+T_0) + \chi(0)[1-(T/T_E)^2]$. The first term of the equation arises from the local magnetic moments of the ions at paramagnetic sites, and the second one is the Pauli paramagnetic susceptibility term which is related to the quasiparticle density of states (DOS) at Fermi level[24, 25]. The fitting yields $c$ = 0.0013 emu K/mol Oe, $T_0$ = 0.8 K, $\chi(0)$ = 0.0006 emu/mol Oe, $T_E$ = 1295.7 K. As for $c = \mu_0\mu_{eff}^2/3k_B$, the effective magnetic moment per Co can be estimated to be around $0.10\mu_B$. In Fig. 2b, we present the specific heat coefficient $C/T$ vs. $T^2$. As shown in the inset, in the low temperature region (T < 10 K), we fit the experimental data with the Debye model $C/T = \gamma_n + \beta T^2$ and $\beta = 12\pi^4 N_A Z k_B T^2/5\Theta_D^3$. Using the value of Avogadro constant $N_A = 6.02 \times 10^{23}$ $mol^{-1}$, Boltzmann constant $k_B = 1.38 \times 10^{-23}$ J/K, Z = 3 (the number of atoms in a primitive cell), the Debye temperature $\Theta_D$ is estimated to be 248.7 K. In contrast to the Debye model fitting as shown by the red line here, there is a slight but clear upturn for the data in the low temperature limit. This upturn may be induced by the Schottky anomaly or the correlation effect. However its smooth temperature dependent evolution together with the large Wilson ratio may exclude the former case. The low temperature upturning of specific heat coefficient and deviation from the Debye model fitting suggests the presence of correlation effect. A polynomial fitting to the data in the low temperature limit gives $\gamma_n$ = 9.75 mJ/mol K$^2$.

**Temperature dependence of resistivity under high pressure.** The temperature dependence of resistivity for ScZrCo$_{1-\delta}$ at various pressures is presented in Fig. 3a. At ambient pressure, the resistivity curve exhibits a rising upon lowering temperature in the whole temperature region from 1.9 K to 300 K, which reveals a bad metal or weak semiconducting behavior. The temperature dependence of resistivity cannot be fitted with the model $\rho = \rho_0 \exp[-k_B T/E_g]$ with a band gap $E_g$. By applying pressure, the resistivity reduces gradually, and the up-going behavior at ambient pressure is suppressed, transforming to a metallic one. As the pressure reaches 19.05 GPa, the superconductivity shows up above 1.9 K with a small but sharp upturning in low temperature region, which may result from the non-standard four-point measurement and the superconducting current passes through a zig-zag way. This upturn can also be suppressed with increasing magnetic field. For this reason we regard the onset transition temperature as the point where this upturn starts. As we increase the pressure further, the superconducting transition temperature gets increased. At 36.05 GPa, the highest pressure we can reach in this run of experiment, superconductivity with an onset transition temperature $T_c$ = 3.6 K has been observed, and meanwhile the zero resistivity is also observed at a lower temperature. Fig. 3b shows the normalized resistivity. As we can see, the value of residual-resistivity-ratio (*RRR*) drops gradually with the pressure below 24.51 GPa. Then, the *RRR* ≈ 1.26 keeps almost unchanged with further increasing the pressure up to 36.05GPa. The inset of Fig. 3b shows the enlarged view of the low temperature resistivity at 16.26 GPa and 36.05 GPa.

**Phase diagram for $T_c$-P and $\rho_{4K}/\rho_{300K}$-P**. To determine the superconducting critical transition temperature, we have adopted different criterions: the onset superconducting transition temperature $T_c^{onset}$ is determined from the very beginning of the deviating point of the normal state resistivity curve; the critical temperatures with 50%$\rho_{4K}$ and 90%$\rho_{4K}$, where $\rho_{4K}$ is the resistivity at 4K. In Fig.4 we present a phase diagram which reveals the doping dependence of the superconducting transition temperature $T_c$ and the ratio $\rho_{4k}/\rho_{300K}$. As shown in the phase diagram, the value of $\rho_{4k}/\rho_{300K}$ is suppressed gradually when the pressure is increased up to 19.5 GPa. At the pressure of about 19.5 GPa, we observe a deviation of resistivity from the normal state background, which as explained above, is due to the presence of superconductivity. The superconducting transition temperature increases gradually by applying higher pressures and $T_c$ goes up to 3.6 K under a pressure of about 36.05 GPa. The transition temperature is expected to rise further under higher pressures. From Fig.3b, we find another interesting phenomenon that when the weak semiconducting behavior is suppressed, superconductivity starts to emerge. This can easily get a reflection from the phase diagram shown in Fig.4. This set of data manifest that superconductivity shows as a subtle balance between the localization and itinerancy of electrons[7].

**Upper critical field.** In Fig. 5, we present the low temperature resistivity under different magnetic fields up to 4 T at 36.05 GPa. The systematic evolution of resistivity curve under a magnetic field tells no doubt that the drop of resistivity at around 3 K is due to superconductivity. We thus determine the upper critical field $H_{c2}$ versus temperature and display the data in the inset of Fig. 5. Here, we use the criterions of 90%$\rho$(4K) and 50%$\rho$(4K) to determine the value of $H_{c2}(T)$. From the curve with a criterion of 90%$\rho$(4K), we have $-dH_{c2}/dT|_{T_c} = 2.7$ T/K. We further fit the $H_{c2}$ curve with an empirical formula $H_{c2}(T) = H_{c2}(0)[1-(T/T_c)^2]^a[1+(T/T_c)^2]^b$. The fitting curve is shown by the red dashed line. For the criterion of 90%$\rho_{4K}$, The parameters of $a$ and $b$ used in the fitting are 0.74 and -0.68, respectively. The estimated $H_{c2}(0)$ is about 5.14 T. Considering the $T_c(90\%\rho_n)$ is 2.7 K, and the ratio of $H_{c2}(0)/T_c$ = 1.9 T/K.

**Discussion**

**Crystal structure and inversion symmetry broken in $ScZrCo_{1-\delta}$.** From the Rietveld fitting, the space group of $ScZrCo_{1-\delta}$ is determined to be *FD3-MS*, in which no inversion center can be found. Four nearest Co atoms in $ScZrCo_{1-\delta}$ form a regular tetrahedron as shown in Fig. 1a. And five Co tetrahedrons, as four vertexes and one centroid, assemble a bigger tetrahedron. As we know, this kind of tetrahedron has a typical non-centrosymmetric geometric structure. Compounds with non-centrosymmetric structure usually behave in a unique way, especially about their superconducting order parameter. Recently, unconventional superconducting properties were observed in the non-centrosymmetric compounds $CePt_3Si$[26] and $Li_2Pt_3B$[27] for the possible existence of spin-triplet pairing symmetry. It is curious to know whether our present system has the spin triplet pairing or not.

**Correlation effect.** The temperature dependence of magnetization at ambient pressure can be well described by the Curie-Weiss law, yielding an average magnetic moment $\mu_{eff} \approx 0.10\mu_B$/Co. It is very similar to the Na$_{0.7}$CoO$_2$, which also shows a Curie-Weiss law behavior and was argued to be induced by correlation effect[18,19]. Specific heat measurements in ScZrCo$_{1-\delta}$ indicate that the low temperature data deviates from the Debye model and exhibit a slight upturn in the low-temperature limit, which may demonstrate the possible correlation effect in our sample. Usually the Wilson ratio can be used to determine whether there is correlation effect in the system. By definition the Wilson ratio is expressed in the form[28] of $R = [4\pi^2 k_B^2/3(g\mu_B)^2]\chi(0)/\gamma_n$. Here the $k_B$ is Boltzmann constant, $g$ is Lande factor which is 2 for a bare electron, and $\mu_B$ is the Bohr magneton. If $R \approx 1$, the material is considered to be the non-interacting electron gas. If $R$ is less than 2 but larger than 1, it corresponds to an interacting Fermi liquid. A strong electron correlation effect would be expected, if the Wilson ratio $R > 2$. For the present compound ScZrCo$_{1-\delta}$, $R$ is determined to be 4.47, which suggests possible strong electron correlation in the compound. In correlated systems, the localization and itinerancy is subtly balanced and the ground state can vary from a Mott insulator to an itinerant metal[7]. High pressure can tune the properties, which may trigger unconventional superconductivity.

**Bad metal behavior.** At ambient pressure, the resistivity curve behaves in a way of weak semiconducting, which is similar to some alloy materials such as Ti$_{67}$Al$_{33}$

(Ref.29). With increasing pressure, the resistivity curve turns to a metallic behavior, but there is still a slight upturn in low temperature region. Large resistivity scale and absence of saturation in low temperature region are the typical characteristics of a bad metal. There are mainly two criterions to judge whether the resistivity scale is large or not: Ioffe-Regal limit[30] $lk_F^{-1} \approx 1$, or Mooij limit[31] $\rho > 0.15$ mΩ cm. Here $l$ is electron mean free path, $k_F$ is Fermi wave vector. The normal state resistivity of ScZrCo$_{1-\delta}$ at 0 GPa is above 0.88 mΩ cm. Even at the highest pressure 36.05 GPa, it is still larger than the Mooij limit. Therefore we are confident that the system is a typical bad metal caused probably by the correlation effect.

**Superconductivity at high pressure.** By applying high pressure, we have successfully tuned the ground state and induced the superconductivity above 19.5 GPa. The superconductivity emerges just after the rapid drop of normalized resistivity which is shown in Fig. 4. And the superconducting transition temperature $T_c^{onset}$ gets to 3.6 K at 36.05 GPa. At this pressure, we obtain the upper critical field $H_{c2}$ = 5.14 T with the $T_c(90\%\rho_n)$ = 2.72K (see the green dots in the inset of Fig. 5.). In the weak-coupling limit for a BCS superconductor, the upper critical field due to the pair breaking of Pauli limit is $H_P$ [T] = 1.84$T_c$ [K][32], which corresponds to be 5.0 T for the ScZrCo$_{1-\delta}$.

In conclusion, we have discovered a new compound ScZrCo$_{1-\delta}$ which has a cubic structure with the space group *FD3-MS* without a spatial inversion center. The ground state of this material is a bad metal showing a strange weak semiconducting behavior. By applying pressure beyond 19.5 GPa, the system becomes superconductive. A superconducting transition temperature at about 3.6 K can be observed under 36.05 GPa. The upturning of low temperature specific heat coefficient and large Wilson ratio suggest a ground state with correlation effect. This material provides a typical platform for achieving superconductivity by finely tuning the subtle balance between localization and itinerancy of $d$-orbital electrons.

## Methods

**Sample preparation and characterization.** The polycrystalline $ScZrCo_{1-\delta}$ samples were fabricated by the arc-melting method. The raw materials scandium (Alfa Aesar, powder, purity 99.8%), zirconium (Aladdin, powder, purity 99.9%) and cobalt (Alfa Aesar, powder, purity 99.99%) were weighed in the mole proportion 1: 1: 0.9, mixed well, and pressed into a pellet in a glove box filled with argon (water moisture and the oxygen compositions in the glove box were both below 0.1 PPM). The pellets were put on the copper base of the Arc furnace filled with argon atmosphere. To ensure the homogeneity of the samples, we re-melted the pellets by turning up and down at least 5 times. Finally, we got the silver-shining polycrystalline samples. The Bruker D8 Advance diffractometer (Cu-$K_\alpha$ radiation) was used to measure the X-ray diffraction pattern. The energy dispersive X-ray microanalysis spectrum and the SEM photograph of the polycrystalline were obtained by S-3400NII (Hitachi). The SEM measurements were performed at an accelerating voltage of 20 kV and working distance of about 10 mm.

**Transport and magnetic measurement at ambient pressure.** The DC magnetization measurements were performed with a SQUID-VSM-7T (Quantum Design). The specific heat of a high quality $ScZrCo_{1-\delta}$ polycrystalline. To ensure the good thermal conduction, we polish one side of the $ScZrCo_{1-\delta}$ sample well and paste with the thermal conduction grease. The specific heat was measured by the thermal-relaxation method with the PPMS-16T (Quantum Design) which allows us to realize high vacuum atmosphere for the sample and measure its specific heat down to 2 K.

**High pressure resistivity measurements.** The measurements under high pressure were accomplished by using the diamond based system cryoDAC-PPMS (Almax easyLab). The culet size of diamonds is 400 μm. The gasket is made by T301 stainless steel which was insulated by the cubic-BN-epoxy mixture, and a small piece of ruby was used as a pressure manometer. The relation of pressure and ruby R(1) fluorescent peak is calculated by the equation $P(GPa) = 380.8[(\Delta\lambda/\lambda_0+1)^5-1]$ [33]. Here $\lambda_0$ is the position of ruby R(1) fluorescent peak at ambient pressure. And $\Delta\lambda = \lambda - \lambda_0$ ($\lambda$ is the position of ruby R(1) fluorescent peak at the pressure *P*.) The hole size of the gasket is 200 μm. The pressure transmitting medium was the $ScZrCo_{1-\delta}$ polycrystalline powder itself. Then we used the PPMS-9T (Quantum Design) to measure the temperature-dependent resistivity.

**Acknowledgements**

We acknowledge the useful discussions with Xiangang Wang and Igor Mazin. This work was supported by the Ministry of Science and Technology of China (Grant No. 2016YFA0300401, 2016YFA0401700), and the National Science Foundation China (NSFC) with the projects: A0402/11534005, A0402/11674164, E0211/51302133.

**Author contributions**

The high pressure resistivity measurements were finished by EYW and JS. The samples were synthesized by JS and XYZ. The specific heat measurements were finished by GYC, and magnetization at ambient pressure measurements were finished by EYW, JS and HL. The data analysis was done by EYW, SJ, XYZ and HHW. HHW coordinated the whole work. EYW, XYZ and HHW wrote the manuscript. All authors have discussed the results and the interpretation.

**Competing financial interests**

The authors declare that they have no competing financial interests.

*Correspondence and requests for materials should be addressed to

zhuxiyu@nju.edu.cn, hhwen@nju.edu.cn.

**Figures and legends**

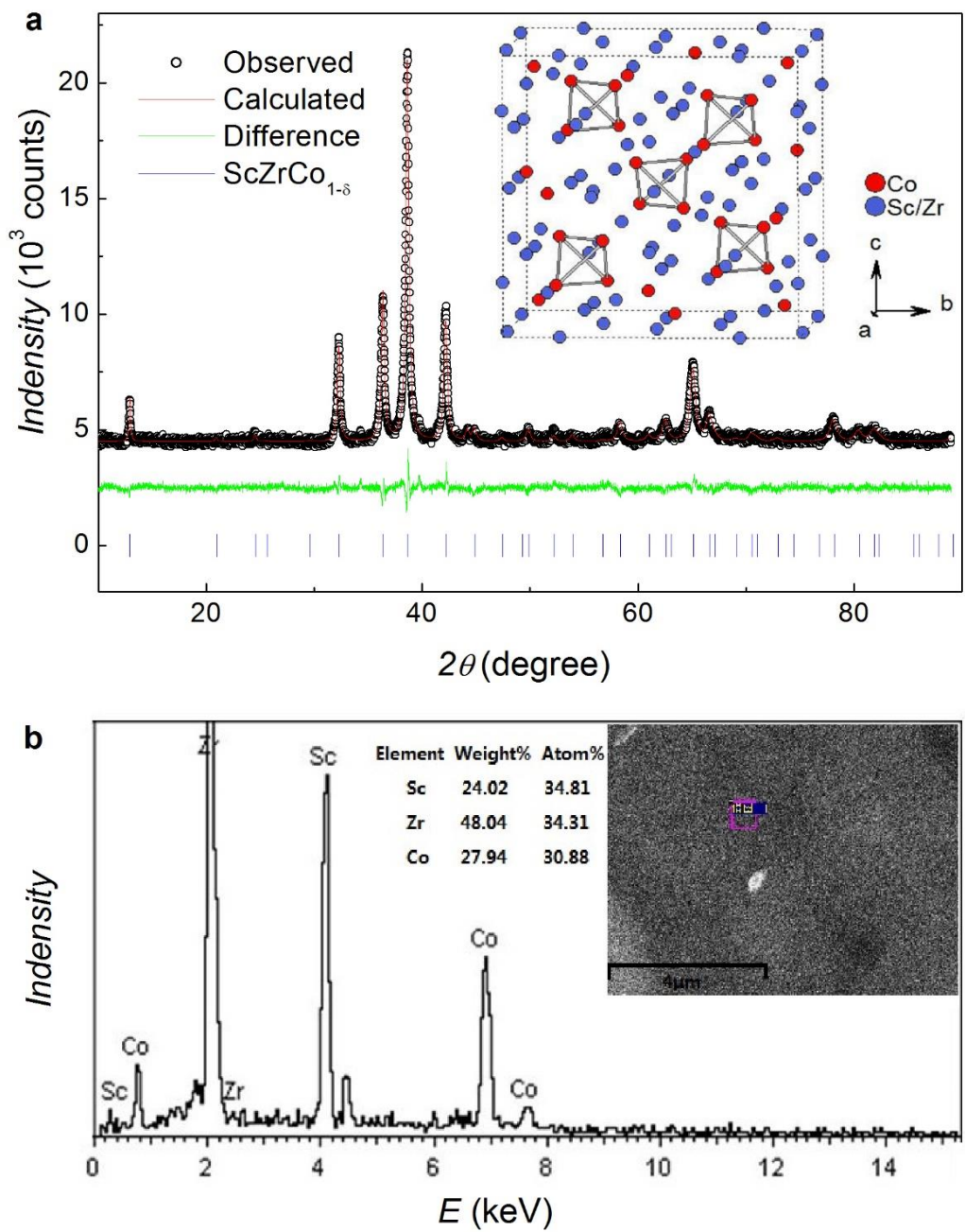

**Figure 1: The XRD and EDS patterns of ScZrCo$_{1-\delta}$ polycrystalline. (a)** X-ray diffraction patterns and Rietveld fit for the ScZrCo$_{1-\delta}$ polycrystalline sample. The inset shows the schematic structure. **(b)** Energy dispersive X-ray microanalysis spectrum taken on a ScZrCo$_{1-\delta}$ polycrystalline. The inset shows the SEM photograph of the sample on which we detect the element composition of the purple line marked area.

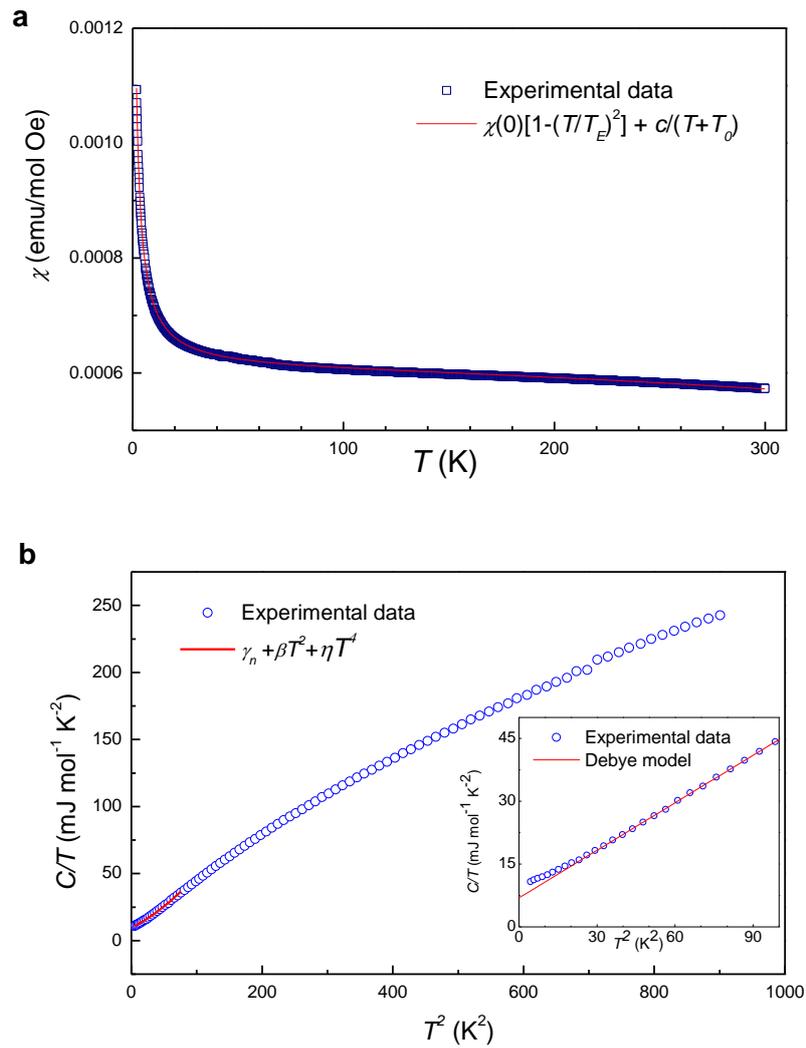

**Figure 2: Magnetic susceptibility and specific heat at ambient pressure. (a)** Temperature dependence of the magnetic susceptibility measured at 1 T. The red line is the fitting curve by the Curie-Weiss law. **(b)** The raw data of specific heat coefficient $C/T$ vs. $T^2$, together with the fitting curve (red line) in the low temperature region. The inset shows the detail of the low temperature region. The red line in the inset presents the Debye model fit, an up-turning of C/T is clearly observed.

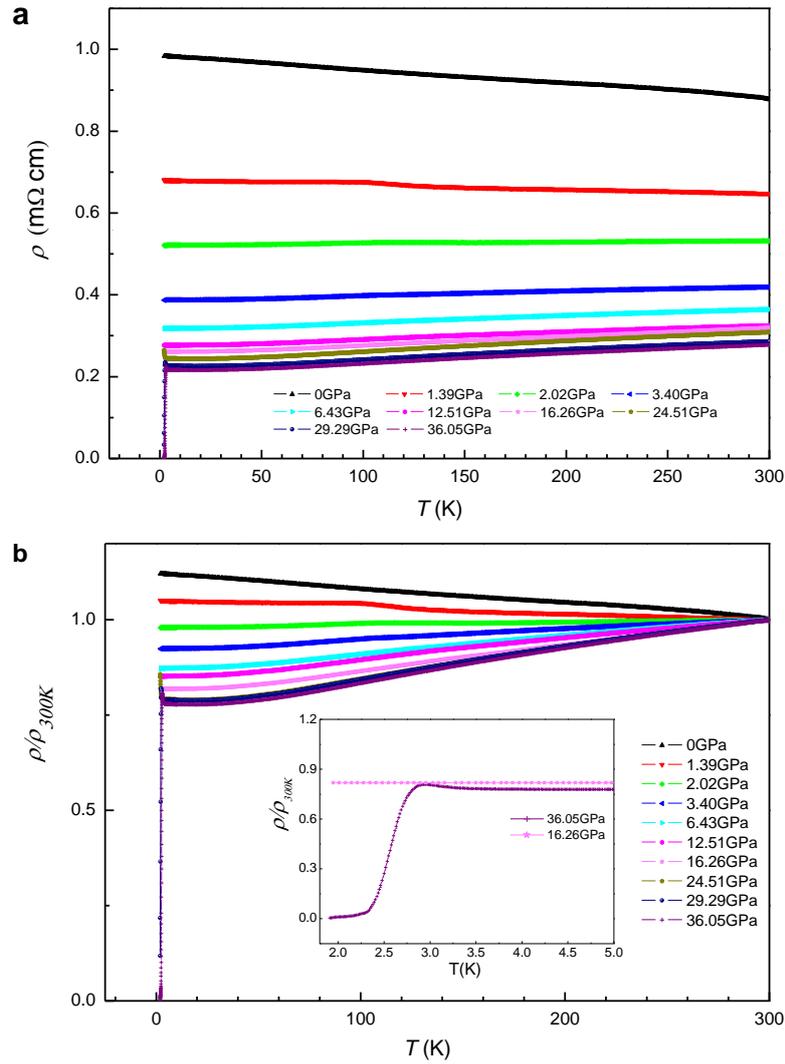

**Figure 3: Temperature dependence of resistivity in ScZrCo$_{1-\delta}$ at high pressure. (a)** Temperature dependence of resistivity for the sample ScZrCo$_{1-\delta}$ in the temperature region from 1.9 K to 300 K at various pressures. **(b)** The normalized resistivity as a function of temperature at various pressures. The inset figure shows the enlarged view of comparison between two curves of resistivity versus temperature with and without superconductivity.

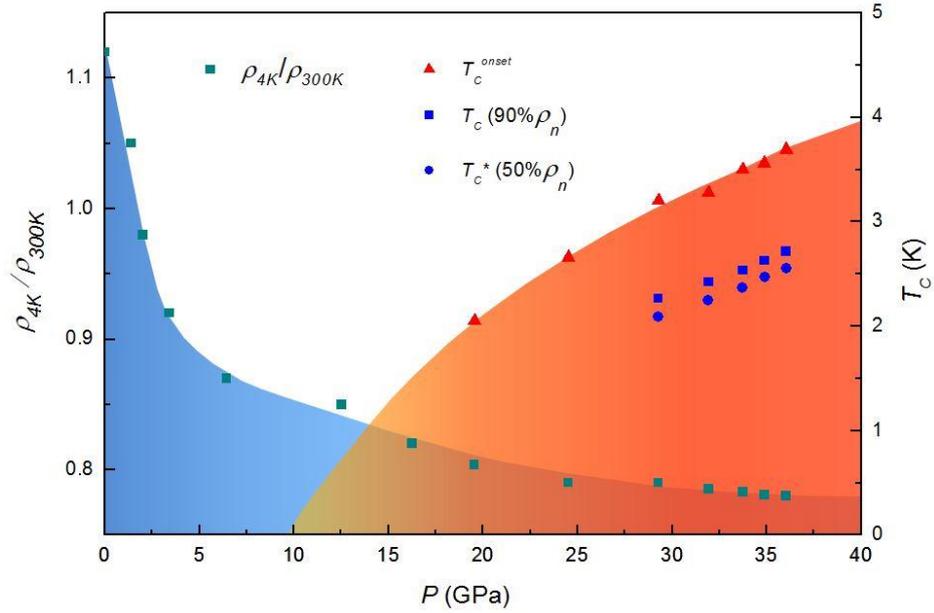

**Figure 4: Phases diagram of ScZrCo$_{1-\delta}$.** Pressure dependence of critical transition temperature $T_c$ and $\rho(4K)/\rho(300K)$ for ScZrCo$_{1-\delta}$. The critical transition temperature is determined by $T_c^{onset}$ (filled red triangles), 90%$\rho_n$ (filled blue squares) and 50%$\rho_n$ (filled blue circles), respectively. The data at 19.54 GPa, 31.95 GPa, 33.76 GPa and 34.9 GPa are not shown in Fig. 3 for clarity. Instead, these data are presented in the Supplementary Information.

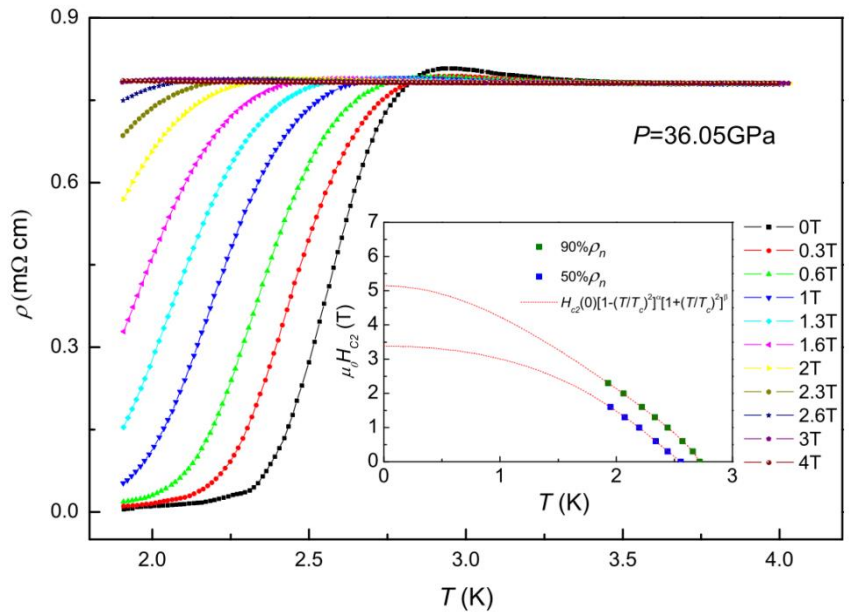

**Figure 5: Resistivity versus temperature under 36.05 GPa and upper critical field.** Temperature dependence of resistivity for the sample ScZrCo$_{1-\delta}$ under the pressure of 36.05 GPa in the temperature region from 1.9 K to 4 K at various magnetic fields. The inset shows the upper critical field $H_{c2}(T)$ determined by the criterions of 90%$\rho_n$ and 50%$\rho_n$, and the fitting curves (see text).

**Table 1: Crystallographic Data of ScZrCo$_{1-\delta}$ at 300 K.**

| compounds | ScZrCo$_{1-\delta}$ | | | |
|---|---|---|---|---|
| space group | $FD3-MS$ | | | |
| $a$ (Å) | 12.2247(9) | | | |
| $V$ (Å) | 1826.939(7) | | | |
| $R_{wp}$ (%) | 1.551 | | | |
| atom | site | occupancy | x | y | z |
| Co | 32e | 1 | 0.912 | 0.912 | 0.912 |
| Sc1 | 16d | 0.5222 | 0.3099 | 0 | 0 |
| Zr1 | 16d | 0.4778 | 0.3099 | 0 | 0 |
| Sc2 | 48f | 0.4334 | 0.125 | 0.125 | 0.125 |
| Zr2 | 48f | 0.5666 | 0.125 | 0.125 | 0.125 |

**Supplementary Information for**

**Pressure Induced Superconductivity in the New Compound ScZrCo$_{1-\delta}$**

Enyu Wang[*], Jing Si[*], Xiyu Zhu[†], Guan-Yu Chen, Hai Lin & Hai-Hu Wen[†]

1. Temperature dependence of resistivity at various pressures. For clarity, these data are not shown in the main text.

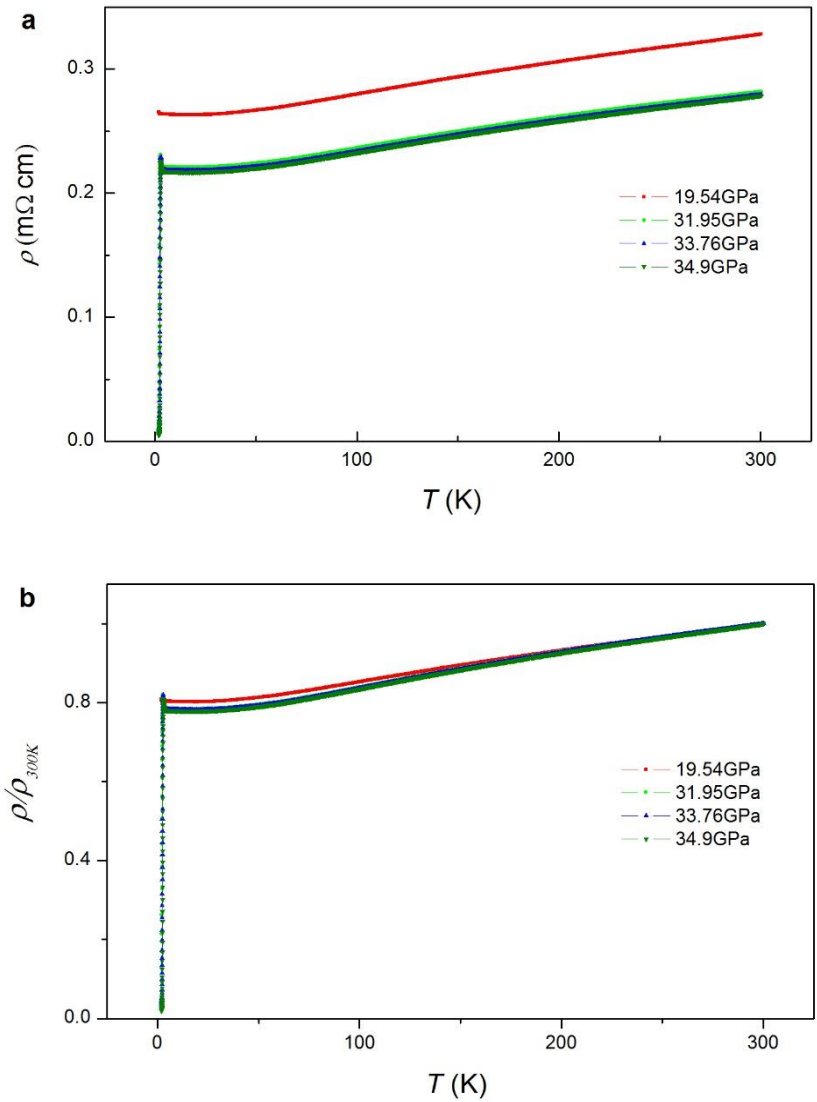

**Supplementary Figure 1: Temperature dependence of resistivity in ScZrCo$_{1-\delta}$ at various pressures.** (a) Temperature dependence of resistivity for the sample ScZrCo$_{1-\delta}$ in the temperature region from 1.9 K to 300 K at various pressures. (b) The normalized resistivity $\rho(T)/\rho_{300K}$ as a function of temperature at various pressures.